\documentclass{article}

\usepackage[english]{babel}

\usepackage[a4paper,top=2cm,bottom=2cm,left=3cm,right=3cm,marginparwidth=1.75cm]{geometry}

\usepackage{amssymb}
\usepackage{siunitx}
\PassOptionsToPackage{hyphens}{url}\usepackage{hyperref}
\usepackage{cleveref}
\usepackage[utf8]{inputenc}
\usepackage[right]{lineno}
\usepackage{csquotes}
\usepackage{booktabs}
\usepackage{longtable}
\usepackage{adjustbox}
\usepackage{array}
\usepackage{url}
\usepackage{titlesec}
\usepackage{authblk}
\usepackage{xcolor} 

\titleformat{\subsection}
  {\mdseries\itshape\large} 
  {\thesubsection}{1em}{} 

\usepackage[english]{babel}
\usepackage[style=authoryear,backend=biber,natbib=true,maxcitenames=2,uniquelist=false]{biblatex}
\DeclareNameAlias{sortname}{family-given}
\DeclareNameAlias{default}{family-given}

\renewbibmacro{in:}{}
\DeclareFieldFormat[article]{title}{\mkbibquote{#1}\addcomma}
\DeclareFieldFormat[book]{title}{\mkbibemph{#1}\addcomma}
\DeclareFieldFormat[bookinbook]{title}{\mkbibemph{#1}\addcomma}
\DeclareFieldFormat[inbook]{title}{\mkbibquote{#1}\addcomma}
\DeclareFieldFormat[incollection]{title}{\mkbibquote{#1}\addcomma}
\DeclareFieldFormat[inproceedings]{title}{\mkbibquote{#1}\addcomma}
\DeclareFieldFormat[manual]{title}{\mkbibemph{#1}\addcomma}
\DeclareFieldFormat[misc]{title}{\mkbibemph{#1}\addcomma}
\DeclareFieldFormat[thesis]{title}{\mkbibemph{#1}\addcomma}
\DeclareFieldFormat[unpublished]{title}{\mkbibquote{#1}\addcomma}
\DeclareFieldFormat[patent]{title}{\mkbibemph{#1}\addcomma}
\DeclareFieldFormat[report]{title}{\mkbibemph{#1}\addcomma}
\DeclareFieldFormat[online]{title}{\mkbibquote{#1}\addcomma}
\DeclareFieldFormat[software]{title}{\mkbibemph{#1}\addcomma}
\DeclareFieldFormat[booklet]{title}{\mkbibemph{#1}\addcomma}
\DeclareFieldFormat[periodical]{title}{\mkbibemph{#1}\addcomma}
\DeclareFieldFormat[standard]{title}{\mkbibemph{#1}\addcomma}

\DeclareFieldFormat[article]{journaltitle}{\iffieldundef{shortjournal}{\mkbibemph{#1}\addcomma}{\mkbibemph{\printfield{shortjournal}}\addcomma}}
\DeclareFieldFormat{volume}{\bibstring{volume}~#1}
\DeclareFieldFormat{number}{\bibstring{number}~#1}

\DefineBibliographyStrings{english}{
  volume = {Vol.},
  number = {No.}
}

\renewbibmacro*{volume+number+eid}{%
  \printfield{volume}%
  \setunit*{\addspace}%
  \printfield{number}%
  \setunit{\addcomma\space}%
  \printfield{eid}}

\renewbibmacro*{journal+issuetitle}{%
  \usebibmacro{journal}%
  \setunit*{\addcomma\space}%
  \usebibmacro{volume+number+eid}%
  \setunit{\addcomma\space}%
  \usebibmacro{issue+date}}

\renewbibmacro*{publisher+location+date}{%
  \printlist{publisher}%
  \iflistundef{location}
    {\setunit*{\addcomma\space}}
    {\setunit*{\addcolon\space}}%
  \printlist{location}%
  \setunit*{\addcomma\space}%
  \usebibmacro{date}}

\DeclareCiteCommand{\cite}[\mkbibparens]
  {\usebibmacro{prenote}}
  {\usebibmacro{citeindex}%
   \usebibmacro{cite}}
  {\multicitedelim}
  {\usebibmacro{postnote}}

\renewbibmacro*{cite:labelyear+extrayear}{%
  \iffieldundef{labelyear}
    {}
    {\printtext[bibhyperref]{%
       \printfield{labelyear}%
       \printfield{extrayear}}}}

\renewbibmacro*{cite:labeldate+extradate}{%
  \iffieldundef{labelyear}
    {}
    {\printtext[bibhyperref]{%
       \printfield{labelyear}%
       \printfield{extradate}}}}

\AtEveryBibitem{
  \clearfield{month}
  \clearfield{day}
  \ifentrytype{book}{
    \clearlist{location}
  }{}
}

\DefineBibliographyStrings{english}{
  andothers = {\textit{et al.},}
}

\DeclareFieldFormat[article]{volume}{\bibstring{jourvol}\addnbspace #1}
\DeclareFieldFormat[article]{number}{\bibstring{number}\addnbspace #1}
\DeclareFieldFormat[article]{volume}{Vol. #1}
\DeclareFieldFormat[article]{number}{No. #1}

\DeclareFieldFormat{url}{\bibstring{available at}\addcolon\space\url{#1}}
\DeclareFieldFormat{urldate}{\mkbibparens{accessed \addspace#1}}

\DeclareFieldFormat{urldate}{%
  \mkbibparens{accessed\space%
    \thefield{urlday}\addspace%
    \mkbibmonth{\thefield{urlmonth}}\addspace%
    \thefield{urlyear}}}

\crefformat{figure}{#2Figure~#1#3}
\Crefformat{figure}{#2Figure~#1#3}
\crefformat{table}{#2Table~#1#3}
\Crefformat{table}{#2Table~#1#3}
\crefformat{section}{#2Section~#1#3}
\Crefformat{section}{#2Section~#1#3}

\author[1*]{Abderrahman Rachidi}
\author[1]{Tarek El Bardouni}
\author[1]{Otman El Hajjaji}
\affil[1]{Radiations and Nuclear Systems Laboratory, University Abdelmalek Essaadi, Faculty of Sciences, Tetouan, Morocco}
\affil[*]{(Corr author)}

\title{$^{7}$Be a Cosmic Window into Atmospheric Dynamics}

\begin{document}
\maketitle

\begin{abstract}
 $^{7}$Be isotope emanating from cosmogenic origins due to high energy cosmic rays, is studied from its production to its detection in the surface, in order to elucidate atmospheric circulation phenomena and analyze the vertical transport of air masses. This can be illustrated briefly by the monsoon model in Kerala in India, where the application of the $^{7}$ Be detection methods in stations in Russia and Australia offered predictions of the debut and retreat of mousson saison in contrast to the meteorological methods.

\end{abstract}

\section{Introduction}
\label{sec:introduction}
Beryllium-7 $^{7}Be$, with a half-life $T_{1/2}=$ 53.2 days, is a radioisotope formed naturally in the stratosphere and upper troposphere \cite{article}. Of cosmogenic origin, it is produced through spallation reactions when high-energy cosmic particles interact with light atmospheric nuclides, such as oxygen $Z = 8$ and nitrogen $Z = 7$.
Regularly detected by the radionuclide detection network of the Comprehensive Nuclear Test-Ban Treaty Organization (CTBTO) on the Earth's surface\cite{longo2019908}, it serves as an excellent proxy for elucidating the physical and meteorological phenomena accompanying atmospheric circulation, as well as the vertical movements of air masses. This is achieved by analyzing the fluctuations in the concentrations of $^{7}Be$ isotopes arriving at the Earth's surface, which are due to both the inverse proportionality between solar activity and galactic and cosmic radiation and the general dynamics of the Hadley and Ferrel atmospheric cells and their convergence zones (HFCZ).

By closely examining the monsoons in the Kerala region of India, it is assumed that they are governed by the seasonal movements of the HFCZ convergence zones and the variation in the circulation of the Hadley cells, which are due to the effects of Earth's rotation and the intensity of exposure to solar activity. For a few months, they are associated with intense rainfall, floods, and storms that can cause agricultural, material, and economic damage. The analysis of $^{7}Be$ isotopes by detection stations in Melbourne, Australia, and Dubna, Russia, allows for the monitoring of the activity of these monsoons, predicting their onset, and forecasting their retreat.
\section{Cosmogenic production and detection of Beryllium-7 isotopes}
\label{sec:method}
\begin{figure}[ht]
 \centering
 \makebox[\textwidth][c]{\includegraphics[width=0.65\textwidth]{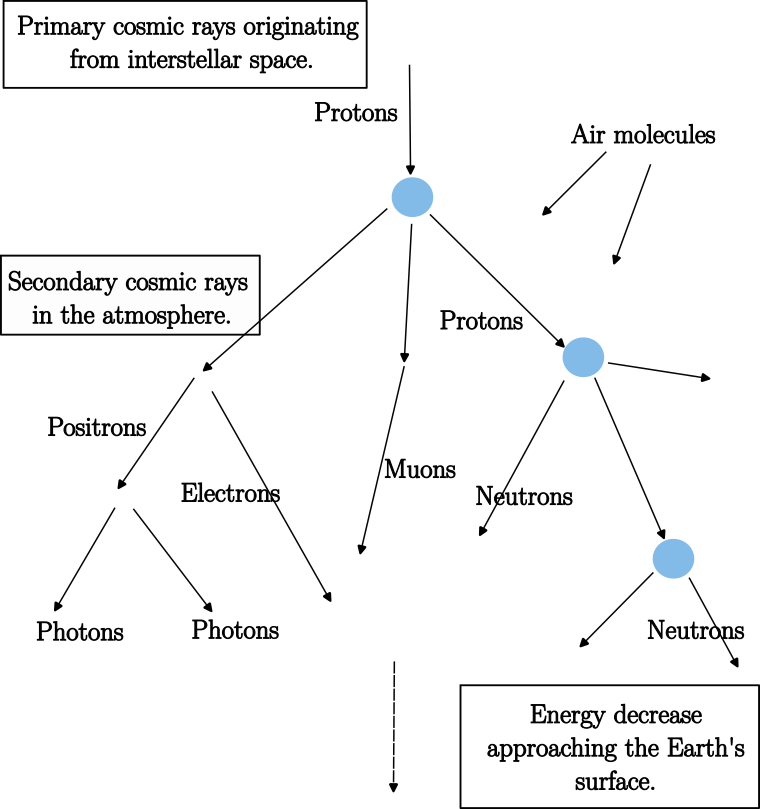}}%
 \caption{ Paths of high-energy particle currents originating from space as they penetrate the atmosphere.}
 \label{f1}
\end{figure}

Beryllium-7 ($^{7}Be$), being a cosmogenic radioisotope, is the result of the Earth's irradiation by high-energy cosmic ray particles (several GeV, even TeV) arriving from interstellar space. These cosmic rays are primarily composed of protons ($89\%$), hydrogen nuclei (the most abundant element in the atmosphere), and helium nuclei ($10\%$). A small proportion ($1\%$) consists of heavier nuclei. These particles collide with atomic nuclei in the upper layers of the atmosphere, producing radionuclides such as $^7Be$, $^{10}Be$, and $^{210}Pb$ through spallation reactions. These collisions also produce a shower of particles (mesons, electrons, muons), but their contribution to nuclear transformations is relatively small compared to nucleons\cite{lal1967},\cite {masarik}\cite{yoshimori20032691}\cref{f1}.
\subsection{Spallation reaction}
These are violent reactions that occur when a target is bombarded with high-energy projectiles, such as protons. These protons shatter the nucleons of the target through inelastic scattering reactions, producing spallation fragments as well as other particles like alpha particles, protons, and neutrons.
\begin{equation}
    p + TNuc\longrightarrow SF_{1}+SF_{2}+...+SF_{p}+ n_{particles}
\end{equation}
\footnotesize {\textcolor{blue}{TNuc}: is the target nucleus} \\
{\footnotesize {\textcolor{blue}{$SF_{i}$} : is a spallation fragment i.}}\\
\normalsize
\\
Although the targets bombarded in spallation reactions are generally nuclei with a high atomic number Z (e.g., lead, bismuth), it is possible to obtain spallation fragments through the bombardment of lighter nuclei (e.g., oxygen, nitrogen, and even carbon) with high-energy protons. Through these processes, $^{7}Be$ can be produced, as seen in reactions $^{16}O(p,5p5n)^{7}Be$ and $^{14}N(p,2p4n)^{7}Be$ \cite{ahmed2007physics}\cite{lusk1967production}.
\cref{f2}\cref{f3}
\subsection{Production Rate of Beryllium-7 in Earth's Atmosphere}
The global average production rate of $^{7}Be$ in the atmosphere can be approximated by the following expression \cite{yoshimori2005production}: 
\begin{equation}
    A\int F(E)(1-sin(\lambda(E)\pi))(1-e^{\frac{-(R(E)-R(E_{0}))}{\Lambda(E)}})dE
\end{equation}
With A representing the probability of $^7Be$ escaping into interplanetary space, F(E) being the sum of the energy spectrum of galactic cosmic rays (GCR) and solar protons, $(1-sin(\lambda(E)\pi))$ being the fraction of the Earth exposed to particles of energy E at the rigidity limit at the geomagnetic latitude $\lambda$, Y(E) being the $^7Be$ yield per spallation reaction, $(1-e^{ \frac{-(R(E)-R(E 0 )}{\Lambda(Y)}})$ being the probability of a particle producing a nuclear interaction below the threshold energy due to ionization loss, R(E) being the energy range of protons with energy E, E0 representing the threshold energy for the production of $^7Be$, and $\Lambda(E)$ being the mean free path for interaction in the atmosphere in $g/cm^{2}$, since it only depends on the amount of matter penetrated, knowing that the oblique atmospheric depth is written \cite{bretz2019zenith} : 
\begin{equation}
    \chi (d)=\int _{d}^{ \infty } (\rho (h(d))d( \textit{d} ))
\end{equation}
With $\rho$ being the atmospheric density as a function of height h and distance along the line of sight d.\\
\begin{figure}[ht]
 \centering
 \makebox[\textwidth][c]{\includegraphics[width=0.65\textwidth]{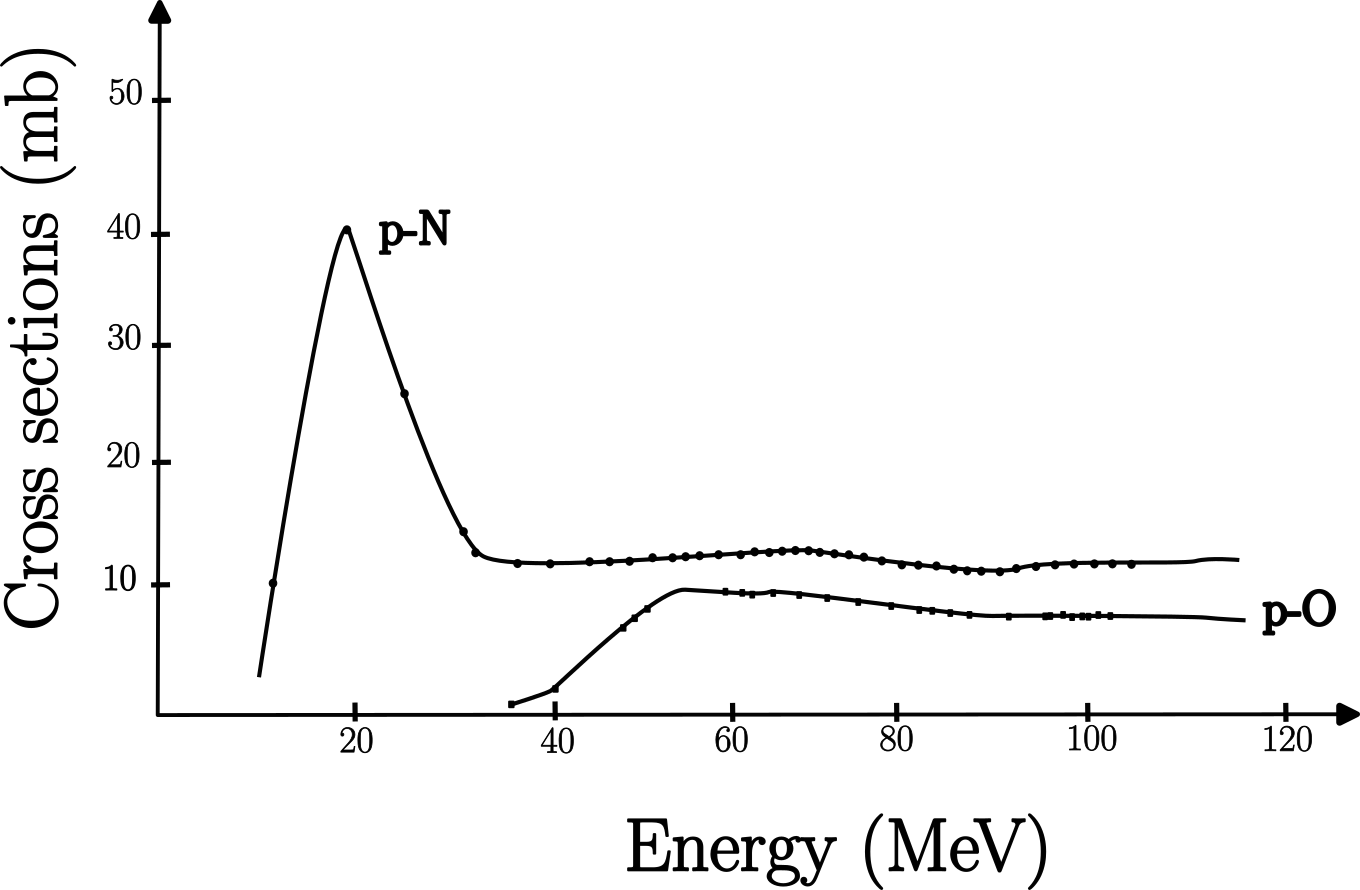}}%
 \caption{ Cross sections of production of $^7Be$ via p-O and p-N reactions\cite{yoshimori2005production}}
 \label{f2}
\end{figure}
\begin{figure}[ht]
 \centering
 \makebox[\textwidth][c]{\includegraphics[width=0.65\textwidth]{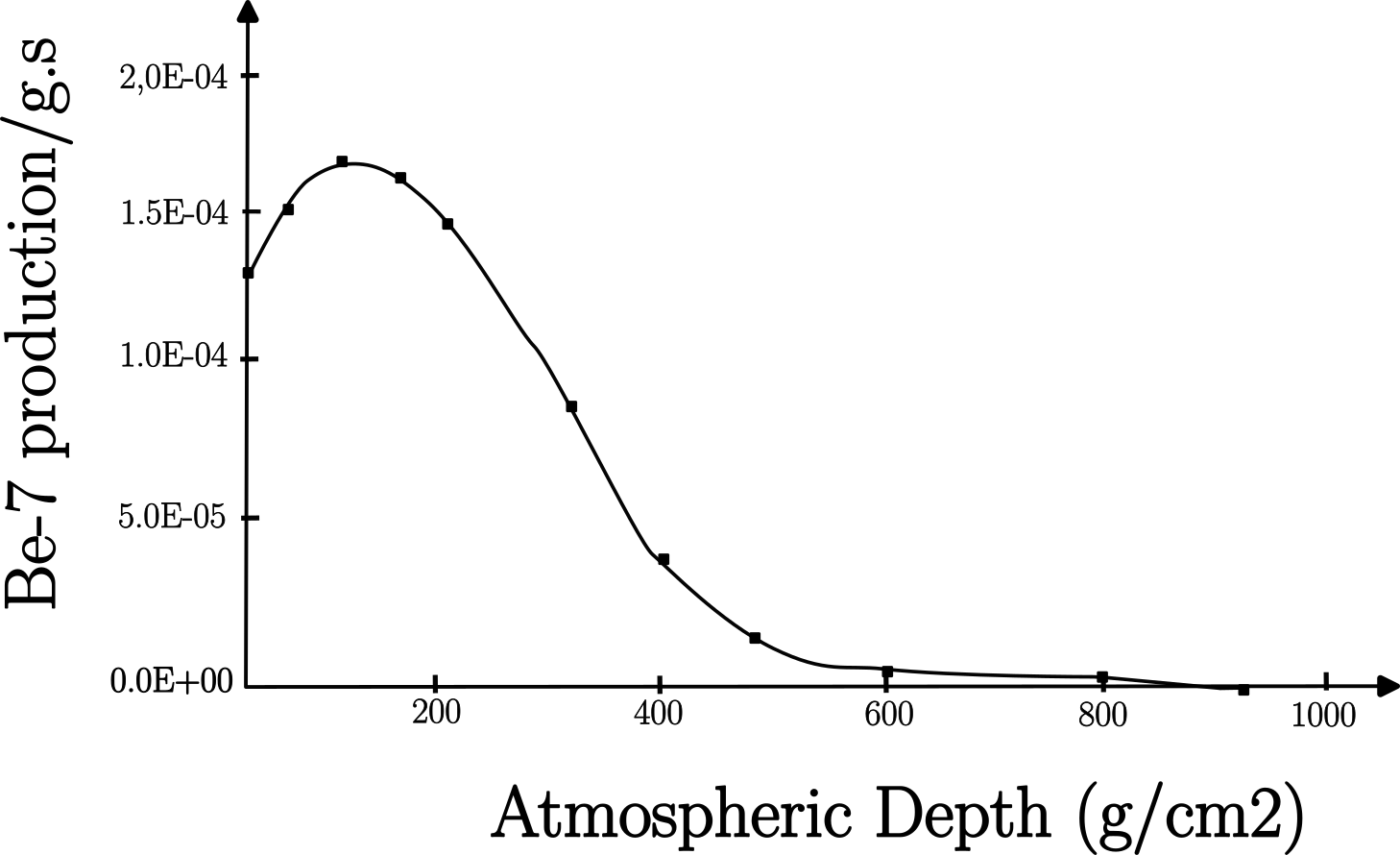}}%
 \caption{ roduction rate of $^7Be/g.s$ as a function of depth\cite{yoshimori2005production}}
 \label{f3}
\end{figure}\\
As shown in \cref{f3}\cref{f2}, approximately $70\%$ of $^7Be$ production occurs in the stratosphere \\ $(< 250 g/cm^2)$ and about $30\%$ in the troposphere $(> 250 g/cm^{2})$.
\subsection{Seasonal variations of Beryllium-7 concentrations.}
\begin{figure}[ht]
 \centering
 \makebox[\textwidth][c]{\includegraphics[width=0.75\textwidth]{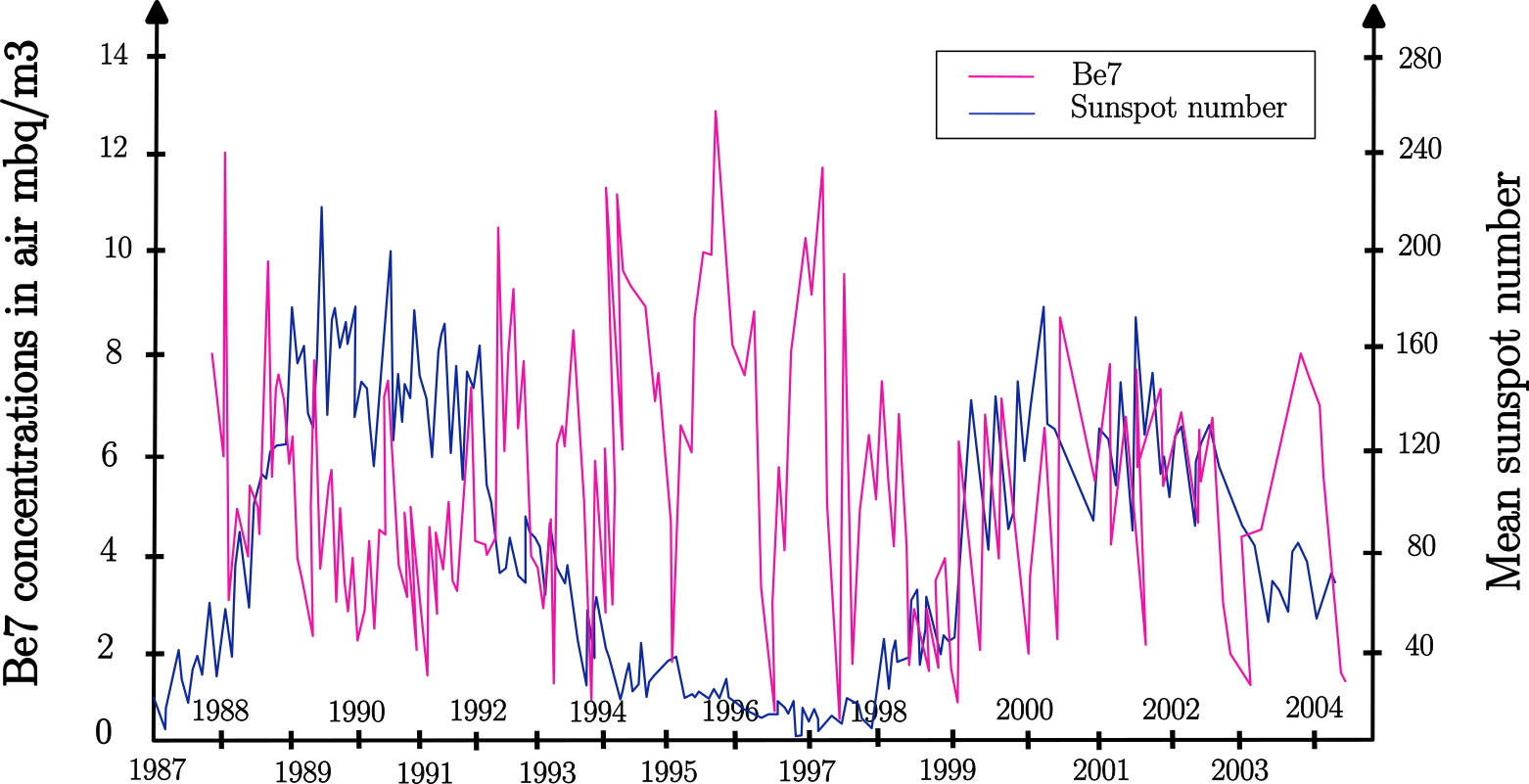}}%
 \caption{$^7Be$ concentrations in the air in function of mean sunspot number between 1987 and 2005\cite{papastefanou2004beryllium}}
 \label{f4}
\end{figure}

\begin{figure}[ht]
 \centering
 \makebox[\textwidth][c]{\includegraphics[width=0.75\textwidth]{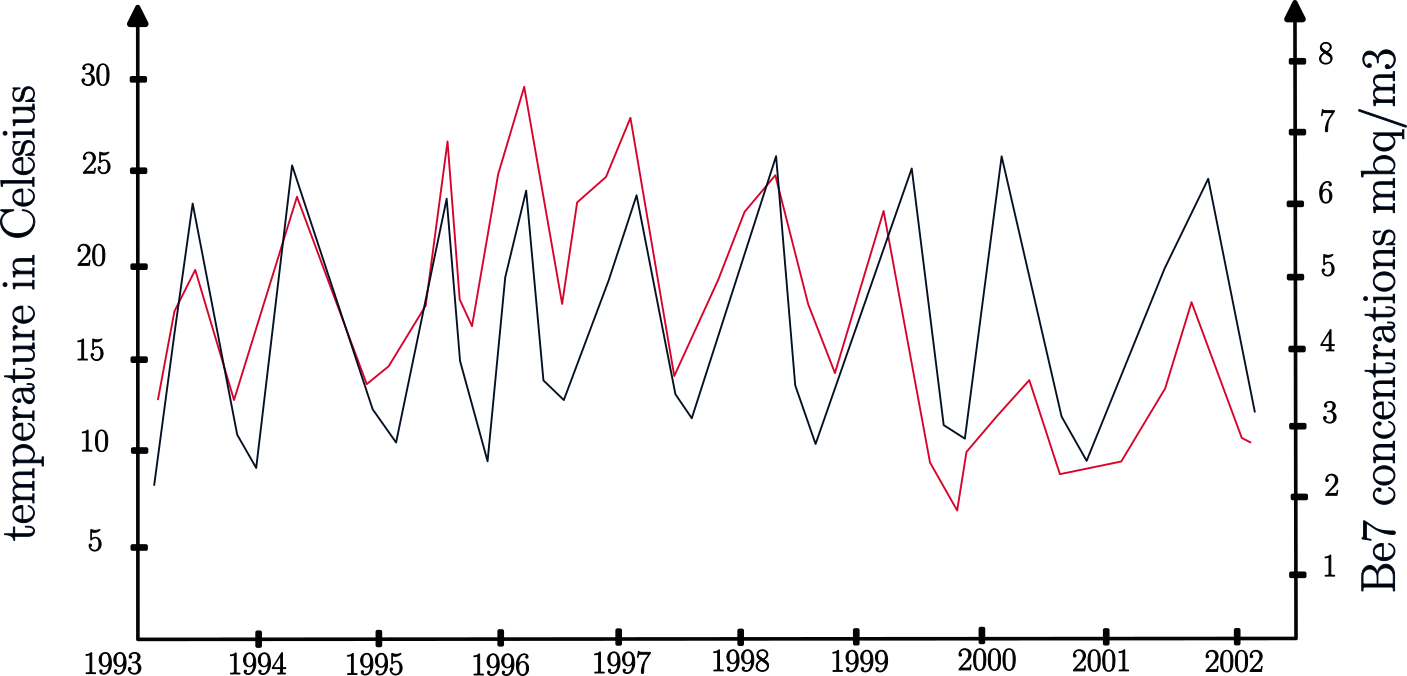}}%
 \caption{$^7Be$ concentrations in function of seasonal temperature in vicnity of the surface in granada Spain between 1993 to 2001\cite{azahra2003seasonal}}
 \label{f5}
\end{figure}

After being produced in the stratosphere, $^7Be$ attaches to submicronic aerosols \cite{azahra2004atmospheric} which are then transported to the troposphere via vertical air mass transport. The production rate of $^7Be$ and its detected concentrations exhibit seasonal variations. The production rate is affected by variations in the cosmic ray flux which are anticorrelated with solar activity\cite{papastefanou2004beryllium}\cref{f4}.
In principle, it is the solar cycles, occurring every 11 years, that are characterized by the intensity of the Sun's magnetic field and sunspots on its surface. This anticorrelation relationship can be interpreted by the interference of solar radiation during the observed peaks of sunspots on the Sun's surface with cosmic rays, eliminating a portion of their passage through the solar system. Regarding the concentrations detected in the vicinity of the Earth's surface, they are influenced by the exchanges between the stratosphere and the troposphere, which experience peaks during the summer season\cref{f5}.
\subsection{Overview of atmospheric circulation patterns}
\begin{figure}[ht]
 \centering
 \makebox[\textwidth][c]{\includegraphics[width=0.75\textwidth]{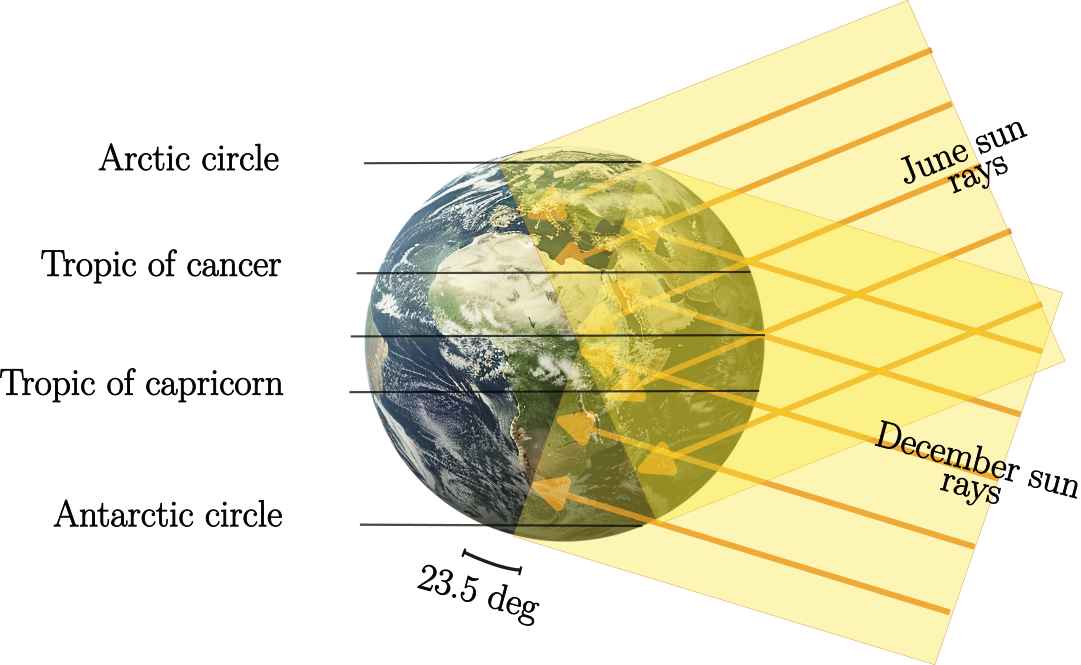}}%
 \caption{Dissimilarity of exposure of the Earth to solar rays}
 \label{f6}
\end{figure}

Due to its spherical symmetry and the tilt of its rotational axis, the Earth is exposed to solar energy in a non-uniform manner. The equatorial region, which divides the Earth into two hemispheres, receives the greatest amount of solar energy compared to the polar regions. \cref{f6}
The Earth's atmosphere plays a role in redistributing this heat imbalance.
Through complex patterns of atmospheric circulation, the air heated in the low latitudes expands and rises up to 10-15 km (the top of the troposphere), leaving behind a low-pressure zone below. Assuming that atmospheric circulation occurs symmetrically in both hemispheres, let's consider the one containing the North Pole. The air heated at the equator moves towards cooler regions due to the principle of heat distribution and collides in the subtropical regions around 30 degrees latitude with the cold air from the pole, creating a zone of turbulence. Where the two winds converge, they descend towards the Earth's surface, creating a high-pressure zone. This current splits into two parts: one that moves towards the North Pole, and the other that forms a cell by rejoining the low pressures at the equator, known as the Hadley cell. Regarding the adjacent current near the Earth's surface that moves northwards, it collides once again with cold air currents from the north, creating a second zone of turbulence. However, this time, limited by the Earth's surface, the air currents rise and split to form a polar cell and another called the Ferrel cell. \cref{f7}\\
\textbf{Hadley Cells}
Named after George Hadley, this atmospheric circulation mechanism is based on three main hypotheses: first, that air circulation within these cells is constant; second, that the thermal wind balance accounts for the circulation; and third, that the conservation of axial angular momentum holds. By definition, the angular momentum of a particle is the cross product of its position relative to the axis of rotation and its momentum. Therefore, as air particles move toward the subtropical regions, they approach the axis of rotation, thus decreasing their distance r. According to the hypothesis of conservation of angular momentum\cite{vallis2017atmospheric}, if r decreases, the quantity of motion $p = m v$ must increase. Since the mass of the particles remains constant, it is the velocity v that increases to compensate for this decrease. Given that these particles are subject to Coriolis forces due to the Earth's rotation, the acceleration of the particles takes an east-west direction.
\\
  \textbf{Air Masses} A large volume of air is defined as an air mass when it possesses relatively uniform physical properties along horizontal planes that sweep across this volume, notably (pressure, temperature, and humidity). By forming through stagnation, it takes on the physical attributes of the region where it is located and, consequently, is classified by two main parameters :
\begin{itemize}
     \item Humidity: Continental air masses are designated as "c" (dry), and maritime air masses as "m".
     \item Region of origin: "T" for tropical, "P" for polar, "M" for monsoon, "A" for Antarctic, "E" for equatorial.
     \begin{figure}[ht]
 \centering
 \makebox[\textwidth][c]{\includegraphics[width=0.65\textwidth]{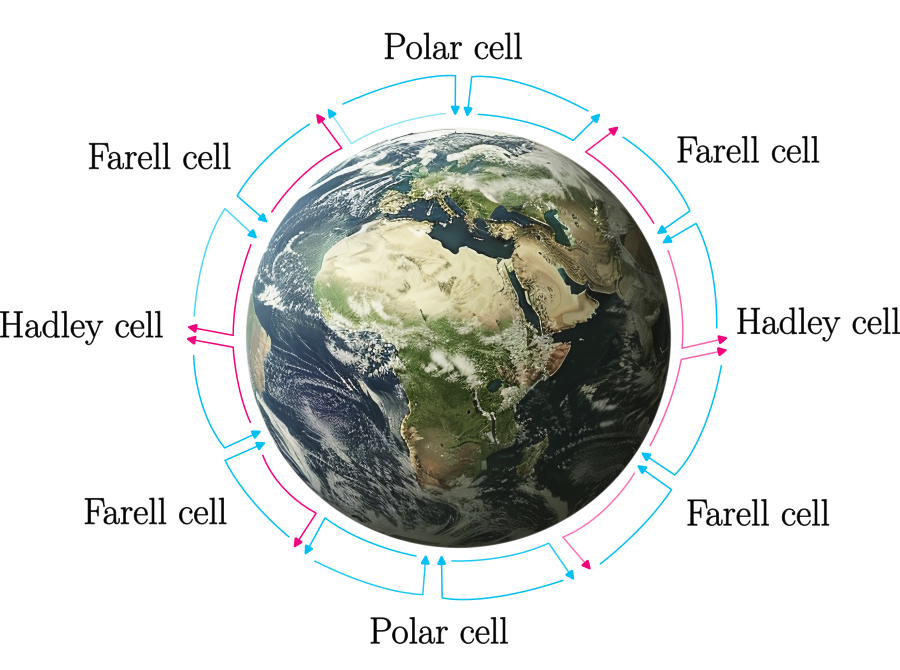}}%
 \caption{Hadley, Farell and Polar cells }
 \label{f7}
\end{figure}
\end{itemize}
A combination of two letters is used to distinguish an air mass. For example, "mT" indicates tropical maritime air masses.
\section{Study of Equatorial Monsoons: A Trans-Approach}
Tropical and subtropical regions, particularly the intertropical zones of Asia, America, Africa, and Australia, experience a weather regime governed by monsoons, characterized by an alternation of dry and wet seasons. These are due to the land/sea distribution as well as the difference in exposure between the hemispheres and the input of solar energy. They are a consequence of the strong thermal contrast that develops between the ocean and continents, creating persistent winds often associated with torrential rains and intense downpours leading to floods. Their unpredictable early or late onset and retreat have a significant impact, with socioeconomic losses estimated at 7 billion dollars in India\cite{vallis2017atmospheric}. On the other hand, produced by spallation reactions, $^{7}Be$ attaches itself to submicrometer aerosols (between 0.08 µm and 2 µm) that are confined in the tropopause. These aerosols undergo deposition processes to find their way to the surface\cite{calec2013depot}. If deposition is carried out by the capture of aerosols by water droplets during cloud formation, it is called wet deposition; through this process, the aerosols undergo downward motions and consequently a washout of the $^{7}Be$ isotopes. Otherwise, if it is governed by atmospheric flow, wind speed,
heat flux and the movement of Hadley and Ferrel cells. It is a dry deposition that is the focus of this study. As previously mentioned, Earth's elliptical heliocentric motion and its axial tilt of 23.4° relative to its orbital plane influence the exposure of the hemispheres to solar radiation. As a result, we observe seasonal shifts in the convergence zones between Hadley and Ferrel cells during the months of December and June. \cref{f6}
\subsection{Trans-Equatorial Approach}
\begin{figure}[ht]
 \centering
 \makebox[\textwidth][c]{\includegraphics[width=1\textwidth]{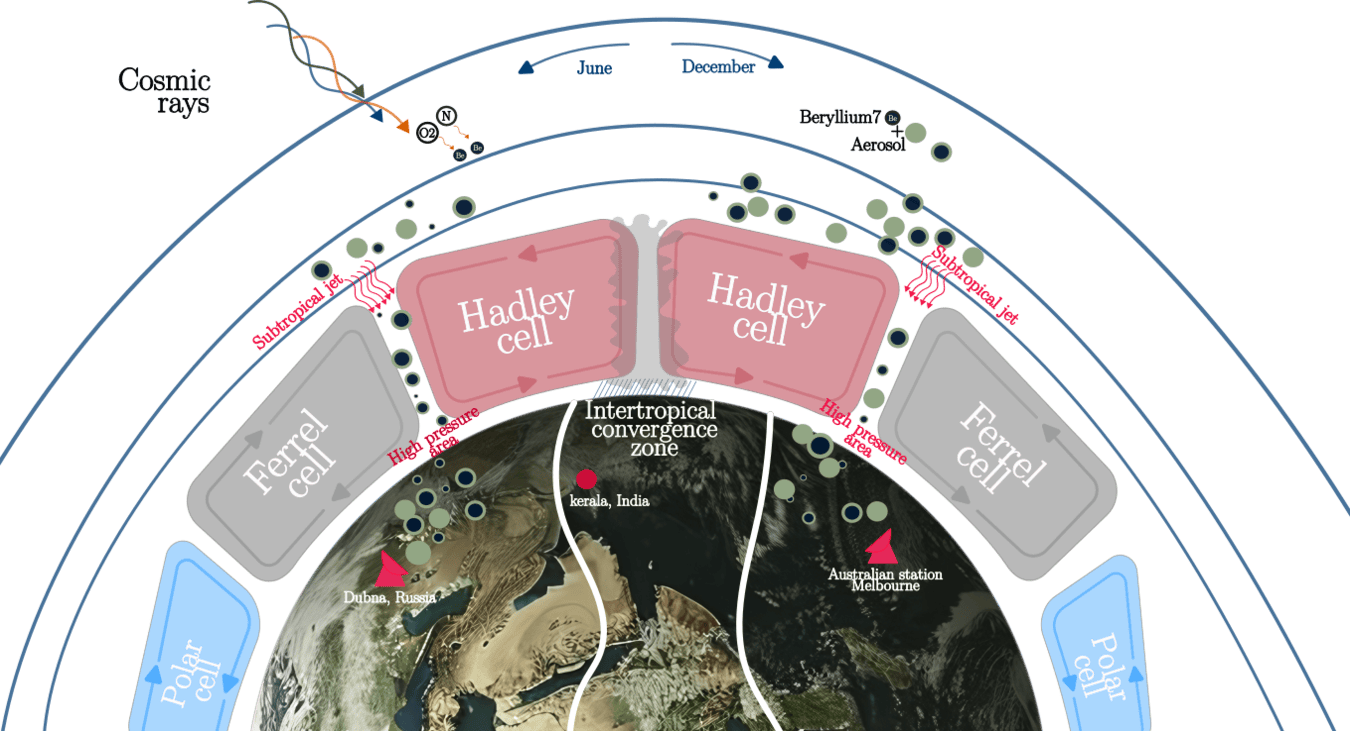}}%
 \caption{movement of the Intertropical Convergence Zone and the processes of production and detection of the isotope $^{7}Be$\cite{terzi2019predict}}
 \label{f8}
\end{figure}

 This approach involves studying the tipping points of monsoons, marking their onset and retreat, by analyzing the seasonal variation of $^{7}Be$ activity. This is done using detection stations located along the two subtropical high-pressure zones in both hemispheres. In the case of Kerala monsoons in India, the chosen stations are Dubna, Russia, located in the northern subtropical region in June, and Melbourne, Australia, located in the southern subtropical region in December. These stations detect a maximum of $^{7}Be$ activity coinciding with the Intertropical Convergence Zone (ITCZ) over Kerala, indicating the onset and retreat of the monsoons, respectively.
 \cref{f8} 
 \\
By analyzing the $^{7}Be$ data, two critical points (cp1 and cp2) can be identified, corresponding to the onset and retreat of the monsoon season. The cp1 predicts the start of the monsoon season with a delay of 53 ± 3 days, and the cp2 predicts its retreat with a delay of 42 ± 7 days \cite{terzi2019predict} \cite{terzi2020radioisotopes}.
\section{Conclusion}
\label{sec:conclusion}
The study of $^{7}Be$, from its cosmic production to its deposition on the surface, proves to be an excellent proxy for tracking the vertical transport of air masses and understanding the mechanisms of atmospheric circulation that govern meteorological regimes. The Indian monsoons serve as a potential model that can be generalized to other regions where monsoons prevail.

\printbibliography

\end{document}